\documentclass[twocolumn,showpacs,preprintnumbers,amsmath,amssymb,superscriptaddress,longbibliography,prl,reprint]{revtex4-1}
\usepackage{color}

\usepackage{graphicx}
\usepackage{dcolumn}
\usepackage{bm}
\usepackage{subfigure}
\usepackage{amssymb}
\usepackage{tabularx}
\usepackage{amsmath}
\usepackage{ulem}
\usepackage{xspace}

\renewcommand{\vec}{\mathbf}
\renewcommand{\emph}{\textit}
\newcommand{\lpa}{\ensuremath{\lambda_\parallel}\xspace}
\newcommand{\lpe}{\ensuremath{\lambda_\perp}\xspace}
\newcommand{\lx}{\ensuremath{\lambda_\times}\xspace}
\newcommand{\leff}{\ensuremath{\lambda^\textrm{eff}}\xspace}
\newcommand{\lzero}{\ensuremath{\lambda_0}\xspace}
\newcommand{\la}{\ensuremath{\lambda}\xspace}
\newcommand{\G}{\ensuremath{\Gamma}\xspace}

\begin{document}

\title{The Effect of Correlations on the Heat Transport in a Magnetized Plasma}

 \author{T. Ott}
 \affiliation{
     Christian-Albrechts-University Kiel, Institute for Theoretical Physics and Astrophysics, Leibnizstra\ss{}e 15, 24098 Kiel, Germany
 } 
 \author{M. Bonitz}
 \affiliation{
     Christian-Albrechts-University Kiel, Institute for Theoretical Physics and Astrophysics, Leibnizstra\ss{}e 15, 24098 Kiel, Germany
 }
 \author{Z.~Donk\'o}
\affiliation{Research Institute for Solid State Physics and Optics, Hungarian Academy of Sciences, P. O. Box 49, H-1525 Budapest, Hungary}

\date{\today}

\begin{abstract}
In a classical ideal plasma, a magnetic field is known to reduce the heat conductivity perpendicular to the field whereas it does not alter the one along the field. Here we show that, in strongly correlated plasmas that are observed at high pressure or/and low temperature, a magnetic field reduces the perpendicular heat transport much less and even {\it enhances} the parallel transport.
These surprising observations are explained by the competition of kinetic, potential and collisional contributions to the heat conductivity. Our results are based on first principle molecular dynamics simulations of a one-component plasma.
\end{abstract}

  \maketitle

Strongly coupled plasmas have become a focal point of plasma research in recent years as several experimental 
setups such as dusty plasmas~\cite{Ivlev2012,Bonitz2010a}, trapped ions~\cite{Jensen2005}, and ultracold plasmas~\cite{Killian2007} 
have become available to study this unusual state of ionized matter in which the potential energy exceeds the kinetic energy. 
Similar plasma conditions are thought to exist, e.g., in white dwarf stars or in the outer layers of neutron stars~\cite{Shapiro1983, Potekhin2010}. 

Often, these plasmas are subject to strong magnetic fields, as is the case in neutron stars, in magnetized target fusion experiments~\cite{Gomez2014}, or 
in recent dusty plasmas experiments~\cite{Thomas2012}. 

The interplay between strong particle interaction and a strong magnetic field gives rise to a rich dynamics~\cite{Dzhumagulova2014}, including complex wave spectra~\cite{Ott2013,Ott2012,Hou2009d}, 
generation of higher harmonics~\cite{Bonitz2010,Ott2011}, and long-lived metastable states~\cite{Ott2013a}. 
The diffusive particle transport has been analyzed in two and three dimensions~\cite{Bernu1981,Ott2011c,Feng2014} and in binary plasmas~\cite{Ott2014} with the conclusion that the self-diffusion coefficient 
is strongly reduced by both the strong coupling and the magnetic field.

Of key importance for plasmas and their applications is their capability to conduct and transport heat.
This plays a crucial role, e.g., in the plasma confinement in modern fusion scenarios~\cite{Gomez2014} or for the 
cooling rate of neutron stars~\cite{Pethick1992}. Heat transport is well understood in magnetized weakly coupled plasmas since the pioneering 
work of Braginskii~\cite{Braginskii1965}, and was studied in strongly coupled unmagnetized systems, e.g., in dusty plasma experiments~\cite{Nunomura2005a,Fortov2007,Nosenko2008}, by kinetic theory~\cite{Suttorp1985}, and by computer simulations~\cite{Donko2000,Salin2002,Donko2004,Donk'o2009,Khrustalyov2012,Kudelis2013}. 
However, until now a systematic study of heat transport in {\it magnetized strongly coupled plasmas} has  not been carried out~\footnote{A preliminary investigation of the effect of a magnetic field on heat transport in a charged two-component 
system is found in Ref.~\cite{Mouhat2013}.}.\nocite{Mouhat2013} It is the goal of this work to fill this gap. By performing first principles molecular dynamics (MD) simulations of one-component plasmas (OCP) 
we show that the heat conductivity $\lambda$ radically differs from the one in weakly correlated high-temperature plasmas~\cite{Braginskii1965,Balescu1988}: a) 
the reduction of $\lambda_\perp$ (the component perpendicular to ${\bf B}$) with increasing magnetic field strength $B$ is much slower
and $\lambda_\perp$ approaches a non-zero asymptotic value, and b) the parallel component $\lambda_\parallel$ {\it increases} with $B$ instead of remaining constant.

{\it Theory and simulation approach.} In a magnetized system, the energy flux $\vec j$ is related to the temperature gradient via the thermal conductivity tensor as $j_\alpha = -\lambda_{\alpha\beta} (\nabla T)_\beta$, where the latter has three independent components (we assume $\vec B \parallel \vec{\hat e_z}$), 

\begin{align}
 \uuline \lambda &= \begin{pmatrix} \lambda_\perp & \lambda_\times & 0 \\ -\lambda_\times & \lambda_\perp & 0 \\ 0 & 0 &\lambda_\parallel \end{pmatrix}.
\label{eq:lambda}
\end{align}
The components \lpa and \lpe describe field-parallel and cross-field heat transport, respectively, and converge to the scalar heat conductivity $\la=\lzero$ as $B\rightarrow 0$. The  off-diagonal term \lx is the analog of the Hall effect (the so-called Righi-Leduc effect~\cite{Delves1965}) and vanishes for $B \to 0$. 

A microscopic approach to $\lambda_{\alpha\beta}$ is provided by the Irving-Kirkwood expression~\cite{Irving1950,Hansen2006} for the heat flux,
\begin{align}
 j_\alpha &= \sum_{i=1}^N v_{i\alpha} \bigg [ \frac{1}{2} m \vert \mathbf v_i\vert^2 + \frac{1}{2}\sum_{j\neq i}^N \phi(r_{ij})\bigg] \nonumber \\
 &\phantom{=}- \frac{1}{2}\sum_{i=1}^N\sum_{j\neq i}^N (\mathbf r_i \cdot \mathbf v_i) \frac{\partial \phi(r_{ij})}{\partial r_{ij}}, \label{eq:j_contribs}
\end{align}
which consists of kinetic, potential, and collision contributions (first, second, and last term). 

The Green-Kubo formula then relates $\underline{\underline{\lambda}}$ to the autocorrelation function~\cite{Evans1990},
\begin{equation}
 \lambda_{\alpha\beta} = \lim_{\tau\rightarrow\infty}\frac{1}{Vk_BT^2}\int_0^\tau \langle j_\alpha(t) j_\beta(0)\rangle dt, \label{eq:gk}
\end{equation}
where the integral is over the energy autocorrelation function. 
According to Eq.~\eqref{eq:j_contribs}, each component of $\lambda_{\alpha\beta}$ is the sum of three direct and three cross-correlation 
terms. 

To compute $\lambda_{\alpha\beta}$ for a typical strongly coupled OCP, we use a screened Coulomb (Yukawa) potential $\phi(r)=Q^2/r\times \exp{(-{\tilde \kappa} r)}$ 
for the pair interaction with the inverse screening length $\tilde \kappa$ and apply a homogeneous 
magnetic field. The system is then completely characterized by three dimensionless parameters: the
coupling parameter $\Gamma=Q^2/(k_BTa)$, the screening parameter $\kappa=a {\tilde \kappa}$, and the normalized magnetic field strength $\beta=\omega_c/\omega_p\propto B$, where $a=\left[3/\left (4 n \pi \right)  \right]^{1/3}$ is the Wigner-Seitz radius, 
$\omega_p= [ 4\pi Q^2 n/m]^{1/2}$ is the plasma frequency, and $\omega_c=qB/(mc)$ is the cyclotron frequency ($Q, T$, and $c$ are the charge, temperature,
and speed of light, respectively). For later use we also define $\alpha = \tau_{col}\omega_p$, where $\tau_{col}$ is the collision time.

We performed extensive MD simulations for magnetic fields~\cite{Spreiter1999,Chin2008}, using 
 $N=8192$ particles in a cubic box with periodic boundary conditions and a value of $\kappa=2$ which is representative for a
strongly coupled OCP. 
Data collection begins after an equilibration period and takes place under microcanonic conditions 
for a time of $t\omega_p=2.5\times 10^5$. Since the evaluation of the integral~\eqref{eq:gk} is limited to a finite time $\tau$ in practice, 
its value is determined by averaging Eq.~\eqref{eq:gk} over $\omega_p\tau\in [1000,2000]$~\footnote{Commonly, the integration in Eq.~\eqref{eq:gk} is carried out only 
up to the first zero-crossing of the integrand. However, we find that the correlation function can have significant oscillatory contributions even for $\beta=0$, accounting 
for a quarter of the value of $\lambda$ at intermediate \G. A detailed analysis of this effect will be published elsewhere.}.
For each data point, $50$ separate simulations with 
different initial conditions are averaged for a total measurement time of $\omega_pt=1.25\times 10^7$. We report the standard error of the mean of these simulations; 
unless shown, the error bar for all values is (much) smaller than the symbol size. Data for the heat conductivity are given in units of $nk_B\omega_pa^2$. 

\begin{figure}
\includegraphics{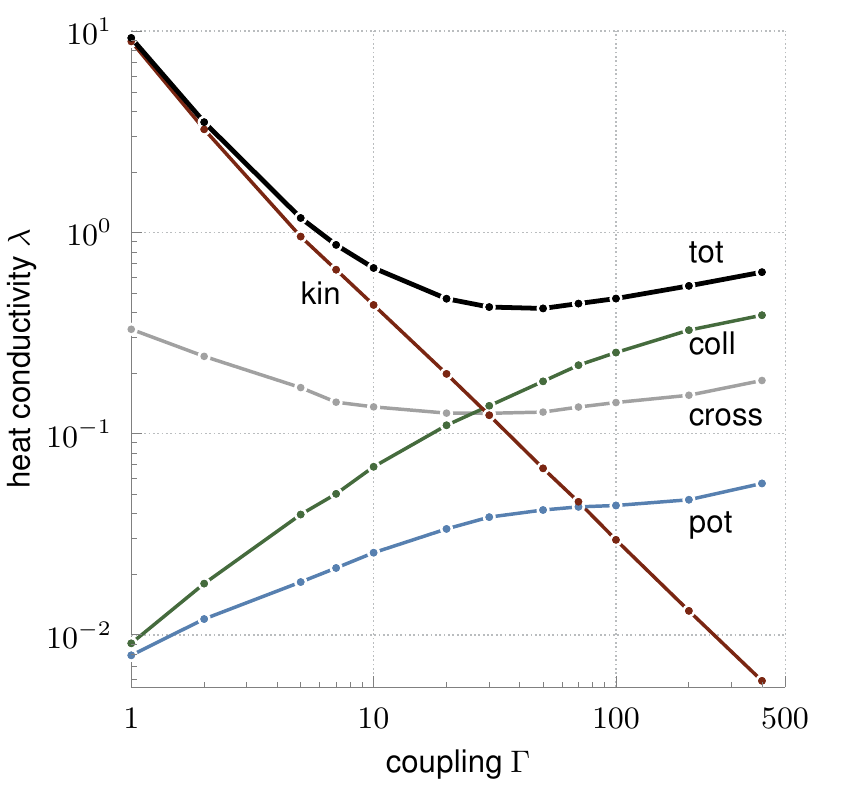} 
\caption{Heat conductivity in an unmagnetized Yukawa system at $\kappa=2$ as a function of $\Gamma$. Shown are the total conductivity and its kinetic, potential, and collisional parts. ``cross'' denotes the sum of the three cross-correlation terms. 
}
\label{fig:k2b0_contribs}
\end{figure}

\emph{Unmagnetized system}---We start with an isotropic, unmagnetized system where $\la_{\alpha\beta}$ 
 is diagonal. Figure~\ref{fig:k2b0_contribs} reveals an interesting non-monotonic dependence of \la on \G, which is one of the hallmarks of strongly coupled systems. The present results represent a significant improvement over previously available 
data and facilitate a more detailed analysis. 

In particular, the decomposition~\eqref{eq:j_contribs} uncovers the origin of this dependency: The kinetic contribution to $\la$ corresponds to the (material) transport of kinetic energy associated with the movement of a particle between collisions. 
It falls off, approximately as $\Gamma^{-3/2}$, since both the mean free path and the thermal energy per particle decrease with $\Gamma$~\footnote{Note that for weakly coupled plasmas ($\Gamma \ll 1$), 
$\lambda$ is proportional to the product of the mean free path and the thermal velocity, i.e., $\lambda_\textrm{wc}\propto\Gamma^{-5/2}$~\cite{Braginskii1965}.}. Likewise, the potential contribution is associated 
with the material transport of potential energy and increases with \G as the average potential energy grows. 
In the region of $\Gamma=30\dots 200$ this growth slows down, since the mean free path decreases with \G, and the system dynamics transform into caged motion. Finally, the collisional part increases with \G up to the crystallization point ($\Gamma_\textrm{c}\approx 420$)
where heat transport is dominated by phonons~\cite{Chugunov2007}. 

\begin{figure}
 \includegraphics{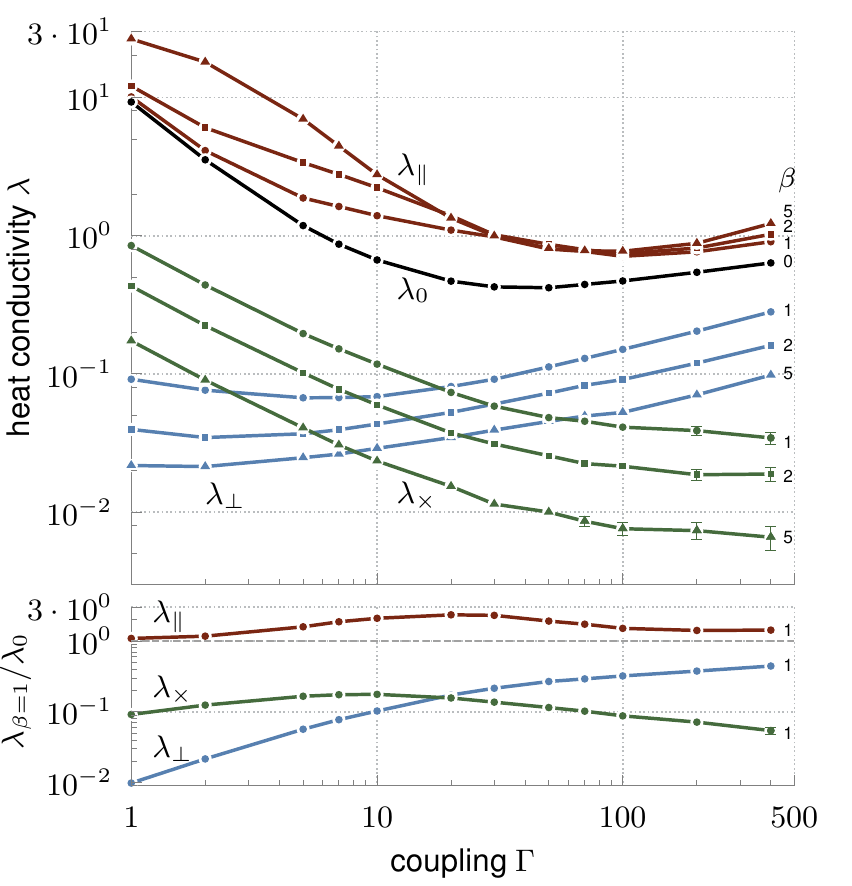}
\caption{Elements of the heat conductivity tensor at different magnetic field strengths as a function of $\Gamma$ (top), and the value at $\beta=1$ relative to the 
field-free value $\lambda_0$ (bottom).  }
\label{fig:k2b0125}
\end{figure}

\begin{figure}
\includegraphics{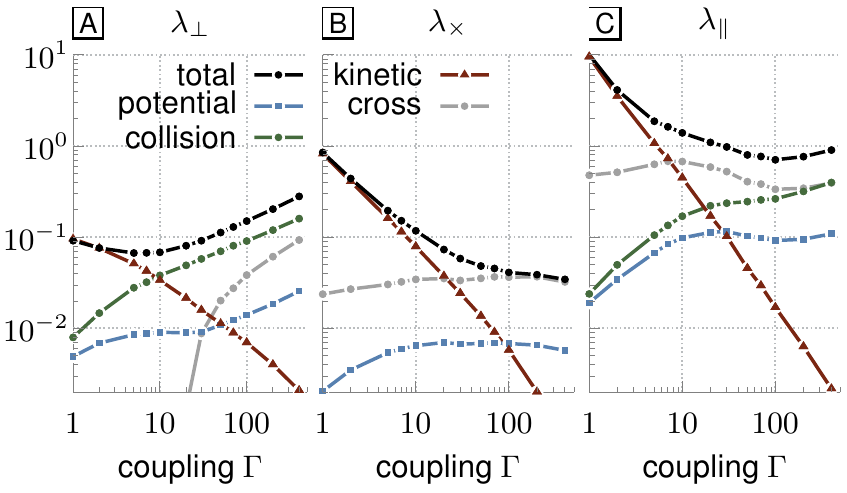} 
\caption{Contributions to the heat conductivities \lpe, \lpa, and \lx at $\beta=1$. ``cross'' denotes the sum of the three cross-correlation terms. 
}
\label{fig:k2b1_contribs}
\end{figure}

\emph{Cross-field heat transport, $\lambda_\perp$}---In the presence of a magnetic field, the \G-dependence of $\lambda$ changes drastically, as is shown in Fig.~\ref{fig:k2b0125}. 
Consider first the field-perpendicular contribution, \lpe. As in a weakly coupled plasma, 
\lpe is reduced by the magnetic field, however, with increasing \G this reduction becomes less pronounced.
This is readily explained by the change of the governing heat transport mechanisms in different coupling regimes: For small \G, material transport of kinetic energy dominates (see Fig.~\ref{fig:k2b0_contribs}) which is greatly reduced perpendicular to {\bf B} by the cyclotron motion of the particles. On the other hand, at large \G, the thermal conductivity is mainly 
due to collisions between particles (Fig.~\ref{fig:k2b0_contribs}) whose frequency and effectiveness are only weakly reduced by the magnetic field. Figure~\ref{fig:k2b1_contribs}a shows how these different effects give 
rise to the observed \G-dependence of \lpe at $\beta=1$. 

\begin{figure}
 \includegraphics{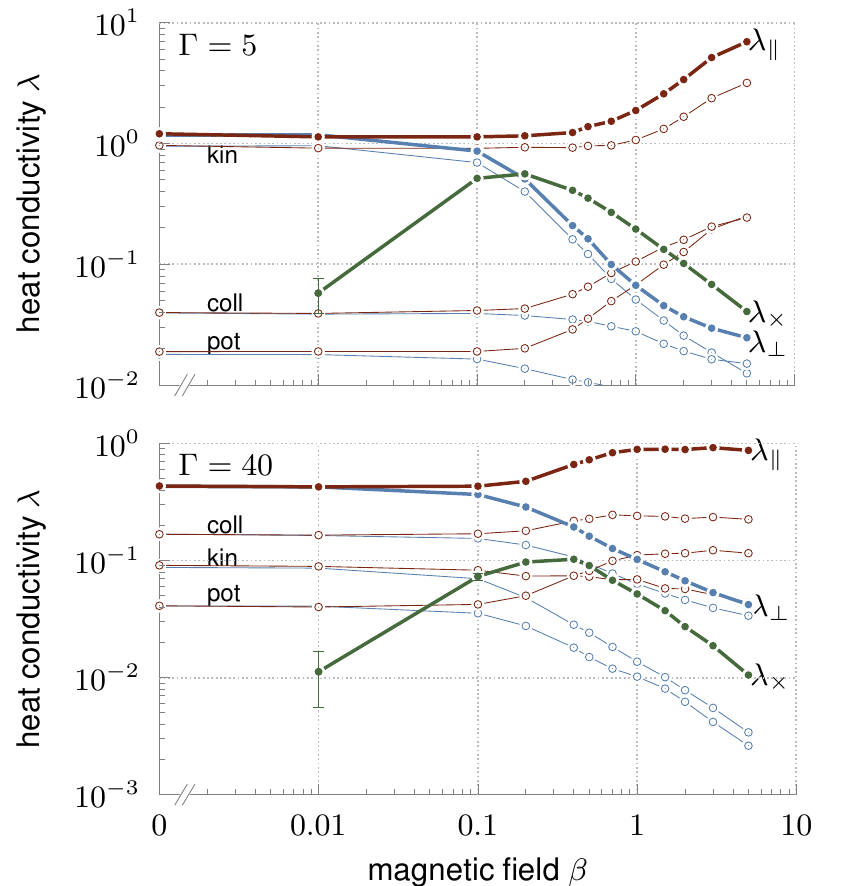} 

\caption{Field dependence of the heat conductivities. Thick lines represent the total values of \lpe, \lpa, and \lx and thin lines 
of the same color show the respective kinetic, potential and collisional contributions. 
}
\label{fig:k2g540}
\end{figure}

Let us now analyze the dependence of \lpe on the magnetic field strength. For weakly coupled systems, both classical transport theory~\cite{Braginskii1965} 
and a hydrodynamic analysis~\cite{Marchetti1987,Marchetti1987a} predict a decay of \lpe scaling as $\lpe~\sim\beta^{-2}$, whereas 
the parallel conductivity is independent of the field strength, $\lpa\sim\beta^0$. 
The corresponding simulation results for strongly correlated systems are shown in Fig.~\ref{fig:k2g540} for the cases $\Gamma=5$ and $\Gamma=40$. 

As observed before, \lpe decreases with $\beta$; however, it does not vanish, but approaches a finite value. The analysis shows that this is due to a residual heat transfer via collisions, see Fig.~\ref{fig:k2g540}: even if particles are unable to  move perpendicular to {\bf B}, heat is still transferred by collective modes of the plasma, in particular, the ordinary shear mode and the upper and lower hybrid modes~\cite{Ott2012}, a mechanism similar to heat transfer via phonons in crystals.

\emph{Thermal Hall effect} \lx---The off-diagonal tensor component (cf. Figs.~\ref{fig:k2b0125}, \ref{fig:k2b1_contribs}b, and \ref{fig:k2g540}) is a special case as it emerges only in the magnetic field. However, an increase of $B$ leads to smaller Larmor radii, which decreases the efficiency of this transport mode. These competing effects result in a non-monotonic behavior with a maximum around $\beta=0.1 \dots 0.5$, cf. Fig.~\ref{fig:k2g540}. With increasing coupling \lx decreases but approaches a finite value, cf.~Fig.~\ref{fig:k2b1_contribs}b, which is explained by the decrease of the  Larmor radius with $\Gamma$ (decrease of temperature)~\footnote{We also note that the collisional contribution to \lx is zero since only the trajectories of the particles  are modified by the magnetic field, but not their interactions.}. 

\emph{Field-parallel transport}, \lpa---Figure~\ref{fig:k2b0125} shows a striking result: heat conduction parallel to ${\bf B}$ is enhanced by the field. This result is surprising, since it is 
in contrast to the behavior of weakly coupled plasmas. But even for strongly coupled plasmas this is an unexpected field effect as it contradicts the behavior 
of the diffusion coefficient parallel to {\bf B} which has been found to decrease monotonically with $B$~\cite{Ott2011c}. 

To clarify the origins of this enhancement, consider again the 
different contributions to \lpa, Fig.~\ref{fig:k2b1_contribs}c. There is a drastic increase of the collisional contribution which is easy to understand: collisions parallel to ${\bf B}$ are becoming more effective with increasing field strength, since particles are unable to avoid one another in lateral direction after the initial approach. This leads to a growing interaction time and, subsequently, an increased energy exchange. This is particularly important for moderate coupling ($\Gamma\lesssim 10$), where infrequent binary collisions dominate, and less so in highly coupled systems, where caged motion already results in highly efficient collisions. 

The potential contribution to \lpa is likewise enhanced by the field. Here, a competition between two processes is observed: On the one hand, the mobility of the particles (i.e., the diffusion coefficient) along ${\bf B}$ is reduced by the field~\cite{Ott2011c}. On the other hand, the reduced interaction with neighboring particles in the cross-field plane enables the particles to retain energy for a longer time, resulting in a net increase of the heat conductivity. 

Similar considerations apply to the kinetic contribution.

The interplay of these processes, together with the varying relative importance of the kinetic, potential, and collisional contributions, work to increase \lpa over \lzero for the whole range of coupling strengths considered. 
Note that the  enhancement of \lpa is weaker for smaller $\Gamma$, and \lpa $\to$ \lzero for $\Gamma \le 1$ (see Fig.~\ref{fig:k2b0125}), confirming consistency of our simulations with the weak coupling limit. 

Consider now the dependence $\lpa(\beta)$. 
As Fig.~\ref{fig:k2g540} shows, \lpa approaches a maximum value as $\beta$ increases. This occurs regardless of whether transport is dominated by material transport ($\Gamma=5$) 
or by collisions ($\Gamma=40$). The physical reasons for this is that both the increase of energy retention and of the efficiency of collisions, which are the driving mechanisms behind the growth in \lpa, 
have upper limits, i.e., complete energy retention and complete collisional energy transfer.

\begin{figure}
\includegraphics{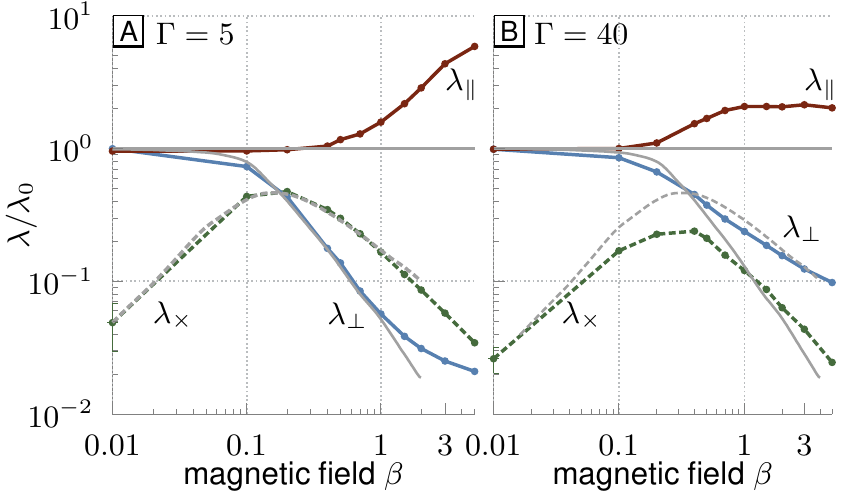} \leavevmode\\
\vspace{0.3cm}
\includegraphics{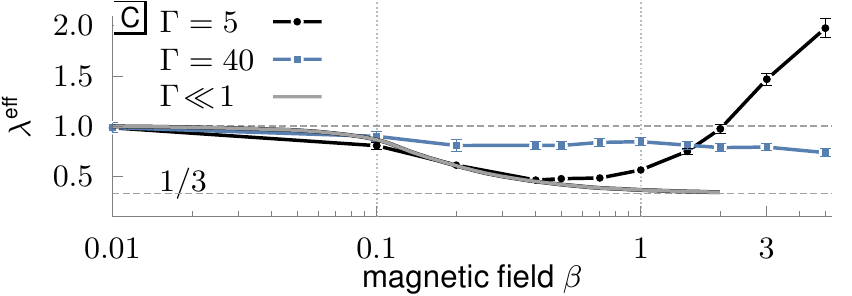} 
\caption{Relative heat conductivities (top) and effective total heat conductivity $\leff=\textrm{Tr}\underline{\underline \lambda} $ (bottom) 
compared to classical transport theory for weak coupling, gray curves~\cite{Balescu1988}. To facilitate the comparison, values of $\alpha=2.5$ 
(a,c) and $\alpha=1.24$ (b) were used (see text). }
\label{fig:balescu}
\end{figure}

\emph{Test of classical transport theory}---Our simulation results allow us to determine the applicability limits of the weak coupling theory of Braginskii~\cite{Braginskii1965,Balescu1988} 
which predicts the scalings $\lpe(\beta)\sim \alpha^{-2}\beta^{-2}$ and $\lx(\beta)\sim \alpha^{-1}\beta^{-1}$ for large $\beta$. A comparison with the strong coupling data at hand shows
semi-quantitative agreement for \lpe and \lx for moderately coupled plasmas, $\Gamma \lesssim 5$, when $\alpha(\Gamma)$ is regarded as a free parameter. Figure~\ref{fig:balescu} 
shows the best fit for $\Gamma=5$ ($\alpha=2.5$) and $\Gamma=40$ ($\alpha=1.24$)~\footnote{The ratio of these values is compatible with the effective Coulomb logarithms for correlated plasmas proposed in Ref.~\cite{Khrapak2013}.}~\nocite{Khrapak2013}
which indicates excellent (poor) agreement in the former (latter) case. At the same time, this extension of weak coupling theory neither captures the finite asymptotics of \lpe for large $\beta$, 
nor can it reproduce the increase of \lpa with $\beta$.

\emph{Summary}---We have presented new high-precision data for the thermal conductivity of unmagnetized and magnetized Yukawa one-component plasmas. The decomposition of
$\lambda$ into the different modes of transport (kinetic, potential, and collisional) has enabled us to elucidate the origin of the well-known non-monotonic dependence of $\la(\G)$ at zero magnetic field 
for the first time. 

At finite magnetic fields, the cross-field thermal transport is reduced, as in weakly coupled plasmas; however, the decay is close to $1/B$ at intermediate field strengths and approaches a finite value at large fields which is due to the collisional mode. On the other hand and, contrary to what one might expect by extrapolating from the weakly coupled case, the field-parallel heat conductivity is enhanced 
by the field in strongly coupled plasmas. This is due either to an enhanced energy retention in systems where kinetic transport dominates or due to a modification of the collision process in systems dominated by this transport mode.

Finally, to understand the implications of our results for strongly coupled plasma applications we consider a heated plasma cube of volume $V$. The energy loss $J_T(\Gamma, \beta)$
through the surface (and, thus, the temperature change) will be proportional to [cf. Eq.~(\ref{eq:lambda})] $\leff = \textrm{Tr}\underline{\underline \lambda} = \lpa + 2 \lpe$. At weak coupling, $J_T$ can be reduced by increasing the magnetic field strength---a standard tool in many plasma applications---to one third of the field-free value (only \lpa remains), see Fig.~\ref{fig:balescu}c. However, in strongly coupled plasmas, our results indicate that using a magnetic field to reduce heat losses is much less effective: for large coupling (cf. curve for $\G=40$) there is almost no reduction whereas, for moderate 
coupling, (cf. curve $\G=5$) losses will be reduced for weak fields, but they will be substantially enhanced in strong fields. Our results have important consequences for the heat transport of strongly coupled magnetized plasmas. One example are the cooling rates of neutron stars and magnetars. The second concerns inertial confinement fusion setups that may, in fact, benefit from the reduction of the heat conductivity and diffusion~\cite{Ott2011c} in strongly coupled plasmas, cf.~Fig.~\ref{fig:k2b0_contribs}. In this case, further substantial reduction of $\lambda$ by a magnetic field is possible but requires the use of asymmetric fuel geometries with a low surface area in field direction.

\begin{acknowledgments}
This work is supported by the Deutsche Forschungsgemeinschaft via SFB-TR 24 project A7, grant shp00006 at the North-German Supercomputing Alliance~(HLRN), and OTKA-K-105476. 
\end{acknowledgments}

\end{document}